# Peculiarities of Hall effect in GaAs/δ<Mn>/GaAs/In$_x$Ga$_{1-x}$As/GaAs ($x \approx 0.2$) heterostructures with high Mn content

M.A. Pankov[1], B.A. Aronzon[1,2,a)], V.V. Rylkov[1,6,b)], A.B. Davydov[1], V.V. Tugushev[1,c)], S. Caprara[3,d)], I.A. Likhachev[1], E.M. Pashaev[1], M.A. Chuev[4], E. Lähderanta[5], A.S. Vedeneev[6], A.S. Bugaev[6]

[1]Russian Research Centre "Kurchatov Institute", Moscow, 123182, Russia

[2]Institute of Applied and Theoretical Electrodynamics, Russian Academy of Sciences, Moscow, 127412, Russia

[3]Dipartimento di Fisica, Università di Roma "La Sapienza", piazzale Aldo Moro, 2 - 00185 Roma, Italy

[4]Institute of Physics and Technology of RAS, Nakhimovskii Avenue 34, 117218 Moscow, Russia

[5]Lappeenranta University of Technology, Box 20, Lappeenranta 53851, Finland

[6]Kotel'nikov Institute of Radio Engineering and Electronics, RAS, Fryazino, Moscow District 141190, Russia

______________________________________________

a)Electronic mail: aronzon@imp.kiae.ru

b)Electronic mail: vvrylkov@mail.ru

c)Electronic mail: tuvictor@mail.ru

d)Electronic mail: sergio.caprara@roma1.infn.it

# ABSTRACT

Transport properties of GaAs/δ<Mn>/GaAs/In$_x$Ga$_{1-x}$As/GaAs structures containing In$_x$Ga$_{1-x}$As ($x \approx 0.2$) quantum well (QW) and Mn delta layer (DL) with relatively high, about one Mn monolayer (ML) content, are studied. In these structures DL is separated from QW by GaAs spacer with the thickness $d_s$ = 2-5 nm. All structures possess a dielectric character of conductivity and demonstrate a maximum in the resistance temperature dependence $R_{xx}(T)$ at the temperature $\approx$ 46K which is usually associated with the Curie temperature $T_C$ of ferromagnetic (FM) transition in DL. However, it is found that the Hall effect concentration of holes $p_H$ in QW does not decrease below $T_C$ as one ordinary expects in similar systems. On the contrary, the dependence $p_H(T)$ experiences a minimum at $T$ = 80-100 K depending on the spacer thickness, then increases at low temperatures more strongly than $d_s$ is smaller and reaches a giant value $p_H$ = (1-2)·10$^{13}$ cm$^{-2}$. Obtained results are interpreted in the terms of magnetic proximity effect of DL on QW, leading to induce spin polarization of the holes in QW. Strong structural and magnetic disorder in DL and QW, leading to the phase segregation in them is taken into consideration. The high $p_H$ value is explained as a result of compensation of the positive sign normal Hall effect component by the negative sign anomalous Hall effect component.

# I. INTRODUCTION

Dilute magnetic semiconductors (DMSs) based on the elements of III-V groups (in particular, GaMnAs and GaMnSb) attract a heightened interest from the point of view of fundamental researches as a new class of materials in which semiconductor and magnetic properties are essentially interconnected [1, 2]. These materials are also of interest for spintronics application due to an important role of spin polarization of carriers in transport and optical phenomena and relatively high Curie temperatures $T_c$ in them [1-3]. Ferromagnet/semiconductor (FM/SC) hybrid structures [3] based on DMSs of III-V groups deserved a special attention due to their quite unusual magnetic and transport properties particularly caused by magnetic proximity effects which exerted a crucial influence in these systems [4-6].

Heterostructures with combined FM and SC thin layers form a separate group of FM/SC hybrid structures. They demonstrate a quite unexpected behavior driven by the magnetic proximity effects due to the charge and spin densities redistribution between FM and SC layers. Last time, the structure with a single internal FM layer deposed near or inside the nonmagnetic SC quantum well (QW) forming the 2D channel of conductivity became a test system to study magnetic proximity effects. In Ref. [7] the photoluminescence (PL) polarization of $Al_{0.4}Ga_{0.6}As/GaAs/Al_{0.4}Ga_{0.6}As$ structures with 0.5ML of FM type of MnAs placed at 5 nm from GaAs QW was investigated. It is important that these structures had a gate which allowed a shifting of holes from QW to FM layer. It was shown that the holes in QW are spin-polarized and that their degree of polarization strongly increases (from 0.4 till 6.3%) with increasing of the "pressing" gate voltage. In the control structures without FM layer the PL polarization was not observed. Induced PL polarization effect was recently observed in other structures, like $GaAs/In_xGa_{1-x}As/GaAs$ ($x \approx 0.2$) containing Mn delta layer (DL) separated from the InGaAs QW by the GaAs spacer with the thickness $d_s$ = 3-5 nm [8]. It is clear that above mentioned results can hardly be explained by a simple tunneling of holes from QW to DL through the spacer, since even under a pressing voltage the depth of such tunneling does not exceed 1 nm (see Fig. 3a from [7]).

Below we present detailed investigation of the temperature and concentration dependences of the transport properties of $GaAs/\delta<Mn>/GaAs/In_xGa_{1-x}As/GaAs$ structures ($x \approx 0.2$) with Mn DL remote from InGaAs QW by GaAs spacer of 2-5 nm width and with high Mn content up to $N_{Mn} \approx$ 1ML. The main goal of our study is to reveal the role of magnetic proximity effects on the Hall effect in these structures. Previously [9], we have found anomalous Hall effect (AHE) in similar structures indicating the spin-polarization of holes in QW. It was also established (see, for example, [10]) that AHE had not been connected with the penetration of Mn atoms inside QW due to their diffusing, but rather explained by an exchange interaction of holes with the remote Mn atoms. Note that the Curie temperatures found in [9, 10] are equal to 30 – 40 K and correspond to the values observed in GaAs/Mn digital alloys [11, 12]. Inserting of DL directly into QW allows to significantly increase the Curie temperature up to $T_C \approx 250$ K [13]. However, the mobility of holes at such location of Mn layer is extremely low (2 - 5 cm$^2$/V·s [13]) and the condition for the formation of 2D holes spectrum is broken [10]. Note also that ferromagnetism in transport properties of In$_{0.75}$Ga$_{0.25}$As/InAs/In$_{0.75}$Ga$_{0.25}$As structures with 2D holes spectrum was observed at temperatures less than 300 mK when internal FM layer had been removed from QW at more than 5 nm [14].

We believe that in structures with relatively high Mn content studied in our work the direct influence of 2D holes in QW on the FM transition in DL is very small, while the peculiarities in the holes transport through the QW are induced by this transition. Note, at the Mn content near 1ML the surface concentration of holes in DL should be $\approx 6 \cdot 10^{14}$ cm$^{-2}$ [13], when the typical concentration of holes in QW does not rise exceed $10^{12}$ cm$^{-2}$. On the other hand, in usually studied GaAs based structures with Mn delta doping the Mn content does not exceed (0.5-0.6) ML [7, 8, 11-13].

As regards the magnetism of the structures under consideration, the mutual influence of QW and FM layer is threefold. First, the quantum magnetic proximity effect (interpenetration of the wave function tails between QW and the FM layer) modifies the effective exchange between the local spins of magnetic ions in FM layer and polarizes the free carrier spins in QW [7]. Second, quantum fluctuations in QW stabilize magnetic order in FM layer, while suppressing the amplitude of the magnetic moment and the transition temperature with respect to those found within the mean-field estimates [15]. Third, electrostatic charge redistribution occurs between QW and the FM layer due to their different density of states and deepness; as a result, modification of the magnetic characteristics of FM layer occurs on purely classical grounds, even without quantum magnetic proximity effect [16].

As it was already mentioned, the long-range character of an interaction between FM layer and QW is difficult to explain only by an ordinary overlap of the ($s$,$p$) –like wave functions of the holes in QW with the $d$-like wave functions of Mn atoms in FM layer, because of a relatively small penetration depth of the hole wave function under the potential barrier formed by the spacer. Recently, however, it was shown that a thin layer of FM metal located into a bulk SC matrix induces quasi 2D spin-polarized collective states, thus a half-metallic type of an electron spectrum of the system is formed [5, 6]. These states (called also confinement states) have however rather extended character along the structure axis of growth because of their small binding energies, so they penetrate deep into the spacer and can reach QW. It can be shown that hybridization of the confinement states with the one electron states in QW leads to their spin polarization, thus a spin splitting of holes induced by FM layer occurs in QW.

Besides the effect of induced spin polarization of holes in QW there exist purely electrostatic effects of charge redistribution between these two sub-systems (for example, the Fermi energy decreasing in QW below FM transition in DL). Indeed, at the one-sided (asymmetric) doping of barriers layers the holes concentration in QW will be determined by a difference of Fermi energies (with respect to vacuum) in Mn DL and QW. So, DL plays a role of «floating gate» with respect to QW, which electric potential is determined by the Fermi level of the system. As a result, the shift down of the Fermi level $\Delta F$ in DL must be accompanied by a decreasing of concentration of holes in QW.

It turned out, that studied structures demonstrate dielectric character of conductivity. Furthermore, the Hall concentration of holes $p_H$ in them does not fall under conditions of FM transition as should be expected and increases with the decreasing of the temperature reaching giant value $p_H \geq \cdot 10^{13}$ cm$^{-2}$ for given type of QW at the holes mobility (10 – 50) cm$^2$/V·s. Meanwhile, in control structures with quasi metallic character of conductivity ($N_{Mn} \approx 0.5$ ML) the drop of $p_H$ is observed at the decreasing of temperature with the subsequent exit on a plateau.

## II. SAMPLES

GaAs/δ<Mn>/GaAs/In$_x$Ga$_{1-x}$As/GaAs structures containing In$_x$Ga$_{1-x}$As ($x \approx 0.2$) QW with thickness $d \approx 10$ nm have been produced by MBE on semi-insulating (001) GaAs substrates at Department of Physics, University of Notre Dame (Fig. 1a). In these structures the DL with content of Mn $N_{Mn} \approx 1$ ML was separated from QW by the spacer of different thicknesses: $d_s = 2$, 3 and 5 nm. While QW and surrounding layers were grown at the temperature ≈600°C, the DL and GaAs cap layer have been grown at 250 °C. Besides, for the compensation effects of QW depletion from the buffer side, the Be acceptor layer with ≈ 30 nm thickness and Be concentration about 10$^{17}$ cm$^{-3}$ has been also placed in the buffer separated from the sample by the spacer of about 5 nm thickness. Prepared structures were subjected by low-temperature annealing at 286 °C in N$_2$ atmosphere during one hour.

Structural properties of discussed systems were investigated by x-ray diffractometry [17]. This method has been previously used for the analysis of GaAs/Mn/GaAs/In$_x$Ga$_{1-x}$As/GaAs systems grown by metal-organic vapor phase epitaxy (MOVPE) at 450 °C and containing a distant broadened Mn layer with much lower Mn concentration [10]. The x-ray diffractometry allows to successfully determine both structural parameters of samples and characteristics of interfaces between layers. The profiles of the lattice parameter $\Delta a/a$ variation in the growth direction were measured and it was shown that for all three samples the interface of QW from the buffer was essentially sharper, than from opposite side of the Mn layer. The x-ray rocking curve and the depth profile $\Delta a/a$ are shown for structure with $d_s$ = 3 nm in Fig. 1b and 1c, respectively. The QW thickness estimated from these data is about 8 nm in agreement with technologically specified 10 nm.

The indium content in QW is in good agreement with that technologically specified ($x$ = 0.2) only for the sample with the biggest spacer thickness ($d_s$ = 5 nm): $x$ = 0.198. For the structures with $d_s$ = 2 and 3 nm the $x$ values calculated from $\Delta a/a$ data turned out to be considerably smaller and diminish with the spacer thickness (0.156 and 0.142 for $d_s$ = 3 and 2 nm, respectively). The probable reason for that is sufficiently high concentration of stacking faults originating on both sides from DL because the Mn content is close to the critical value $N_{Mn}$ ≈ 1 ML [18]. It was shown [18] that at such Mn content a high quality digital GaAs/Mn alloy could not be grown even at high thickness of GaAs spacer (17 ML).

It is known that magnetic order of quasi-2D dimensional structures with a single DL is frequently detected from the study of temperature dependences of anomalous Hall effect and longitudinal resistance, since direct measurements of magnetization in these structures are difficult [13]. This is why in our work we pay attention to the transport properties of GaAs/δ<Mn>/GaAs/In$_x$Ga$_{1-x}$As/GaAs structures which were investigated in the temperature range 5 – 200 K and at magnetic fields up to 2.5 T with the samples of the double Hall cross shape with the width of conducting channel $W$ = (1-1.5) mm and the distance between potential probes $L_p$ = (2.5-3) mm.

To reveal a role of the Mn content, the same measurements were performed for reference GaAs/Mn/GaAs/In$_x$Ga$_{1-x}$As/GaAs structures with $N_{Mn}$ ≈ 0.5 ML and $d_s$ = 3 nm [10]. These structures possessed metallic (not activation) type of conductivity and high holes mobility ($\mu$ ≈ 3000 cm$^2$/V·s at $T$ = 5 K).

### III. RESULTS AND DISCUSSION
#### A. The temperature dependence of resistance

At first, let us consider obtained experimental results on the temperature dependence of the resistance $R_{xx}(T)$. Fig. 2 shows the functions $R_{xx}(T)/R_{xx}$(200 K) for samples 1 – 3 with $N_{Mn}$ ≈ 1 ML and different spacer thicknesses $d_s$ = 2, 3 and 5 nm, respectively (curves 1 – 3). For comparison, the $R_{xx}(T)/R_{xx}$(200 K) dependence of the control sample 4 ($N_{Mn}$ ≈ 0.5 ML and $d_s$ = 3 nm) is also shown (curve 4). Unlike to the $R_{xx}(T)$ dependence for the reference metallic sample, the resistance for the structures 1 – 3 with $N_{Mn}$ ≈ 1 ML sharply increases with a decreasing of temperature, indicating an insulating character of their conductivity. It should be noted that all $R_{xx}(T)$ curves shown in Fig. 2 demonstrate a peculiarity at $T= T_R$ (either a maximum for samples 2-4 or a shoulder for the most insulating sample 1, that is a common property of the III-Mn-V DMSs (see Ref. [2] ). The characteristic temperature $T_R$ is commonly associated with the temperature of FM transition in DMS and widely used to determine the Curie temperature $T_C \approx T_R$ [2,10,13,17]. In our case, for the samples with high Mn content ($N_{Mn} \approx$ 1ML) the temperature of $R_{xx}(T)$ peculiarity does not depend on the spacer thickness $d_s$ within 10% accuracy and is equal to $T_R \approx 46$ K, while in the reference sample with moderate Mn content ($N_{Mn} \approx$ 0.5ML) this temperature is equal to $T_R \approx 30$ K. An insignificant dependence $T_R(d_s)$ for the samples with a high Mn content reflects a relatively weak feedback of the holes in QW on the local spins in DL.

The resistivity $R_{xx}(T)$ is driven by two factors, i.e. the holes concentration $p(T)$ and the holes mobility $\mu(T)$, respectively. Both these factors are sensitive to magnetic ordering and can exhibit significant anomalies near the FM transition point, so it is very difficult to extract the $p(T)$ and $\mu(T)$ peculiarities separately only analyzing the $R_{xx}(T)$ dependence. It is known for example, that near $T_C$ the scattering of holes on critical spin fluctuations leads to significant decrease of $\mu(T)$ [2], i.e. $R_{xx}$ increases with the temperature lowering when $T \geq T_C$. Below $T_C$ spin fluctuations are frozen and the spin dependent part of $\mu(T)$ increases, i.e. $R_{xx}(T)$ decreases at $T \leq T_C$ under these conditions. On the other hand, the holes concentration $p(T)$ can also be dependent on the FM ordering and at first glance, it seems to be natural to explain the increase of $R_{xx}(T)$ below $T_C$ simply by a decrease of $p(T)$ in QW, due to the Fermi level lowering inside DL in the FM regime in DL (see Ref.[16]).

Really, according to calculations in Ref. [16], the FM state in a model of a single delta layer is expected to be strongly spin-polarized like the state of 2D FM half-metal, where the Fermi level significantly decreases below $T_C$. Let us write an evident condition describing the holes redistribution between QW and DL in the simplest form:

$$p = p_0 - C_s(|\Delta F|/e^2) \approx p_0 - (\kappa/4\pi d_s)(|\Delta F|/e^2), \qquad (1)$$

where $p_0$ and $p$ are the holes concentration in QW in the paramagnetic (PM) and FM half-metal regimes, respectively, $C_s$ is capacity of the surface unit between DL and QW, $\Delta F$ is the Fermi level shift in FM half-metal regime, $\kappa$ is permittivity, $e$ is elementary charge. At small spacer thicknesses, $d_s \sim 1$ nm, and at the Fermi level changing $\Delta F \sim -10$ meV, the variation of holes concentration turns out essentially, $\Delta p = p_0 - p \sim 7 \cdot 10^{11}$ cm$^{-2}$. According to Eq.1, at $\Delta F$ increasing in FM half-metal regime the holes concentration $p$ decreases, and for equal $\Delta F$ values the $R_{xx}$ increasing due to the FM transition should be stronger for the thinner spacer, as it is seen from

Fig. 2. However, in our system the situation turns out to be much more complicated as follows from the Hall effect results.

**B. Hall effect measurements**

Fig. 3 shows the dependences of the holes concentration $p_H(T)$ for samples 1 and 3 (with $d_s$ = 2 and 5 nm) obtained from the Hall effect measurements. It is seen that above $T_R$ but at $T < 200$ K, the dependence $p_H(T)$ experiences a minimum at some temperature $T = T_H > T_R$. The value $T_H \approx 110$ K for the sample 1 with $d_s = 2$ nm is higher than $T_H \approx 80$ K for the sample 3 with $d_s = 5$ nm. Besides, the increasing of $p_H(T)$ at $T < T_H$ is more strong in sample 1 than in sample 3. Furthermore, $p_H(T)$ only slightly decreases below $T_R$ and at low temperatures ($T \approx 5$ K) the $p_H$ value turns out to be huge, $p_H \approx (1-2) \cdot 10^{13}$ cm$^{-2}$. Such high value of 2D holes concentration was obtained in heterostructures on the basis of wide gap semiconductors like AlGaN/GaN and AlInN/GaN [20]. Our estimations for heterostructure with one-sided doping (triangle potential well) show that even for a 3D electron spectrum in QW at surface concentration of carriers in it equal to $2 \cdot 10^{13}$ cm$^{-2}$ the barrier exceeding $\approx 0.4$ eV is required. It is not achieved for GaAs/In$_x$Ga$_{1-x}$As/GaAs heterostructures (at $x \leq 0.2$ the barrier does not exceed 0.15 eV) in which the holes concentration in QW at helium temperatures does not exceed $10^{12}$ cm$^{-2}$ [9, 10].

We see a complete unlikelihood in the behavior of $p_H$ and a true concentration of holes $p$ in our structures, since this contradicts to previously discussed $R_{xx}(T)$ dependence. Moreover, it is also impossible to explain obtained values of $p_H$ by the contribution of conductivity inside DL, since the mobility of holes $\mu$ in our case is high enough. In particular, it is about 50 cm$^2$/V·s for the structure with $d_s = 5$ nm (see Fig. 3b), while in GaMnAs films [21] or in GaAs/AlGaAs heterostructures with DL inside the conducting channel [13] the value of $\mu$ does not exceed $\approx 5$ cm$^2$/V·s. On the other hand, in the reference sample ($N_{Mn} \approx 0.5$ ML) the holes concentration $p_H$ decreases at $T < 60$K reaching constant value $p_H \approx 7 \cdot 10^{11}$ cm$^{-2}$ at $T < T_R \approx 30$ K (see Fig. 4), and at the same time the holes mobility increases up to $\approx 3 \cdot 10^3$ cm$^2$/V·s.

Let us come back to the $R_{xx}(T)$ measurements. The inset to Fig. 2 the illustrates the log$R_{xx}(T)$ versus $1/T$ dependences for samples 1 and 3 with $d_s = 2$ and 5 nm. We see that at high temperatures $T > T_0$, where $T_0 \approx 60$K is some characteristic temperature, both samples demonstrate a pronounced activation type of conductivity. However, at temperatures $T < T_0$ the activation energy $\varepsilon_a$ determined as the slope of the log$R_{xx}(1/T)$ dependence sharply decreases. For the most insulating sample 1 the activation energy $\varepsilon_a$ decreases from $\varepsilon_a \approx 16$ meV at $T \geq 90$K to $\varepsilon_a \approx 0.05$ meV at $T \leq 25$ K. Such behavior corresponds to a well-known mechanism of the percolation transport connected with an activation of carriers to the percolation level, which becomes the hopping transport through localized states with temperature decrease [22]. Furthermore, near the transition to the hopping transport the minimum in the $p_H$ (T) dependence (i.e. the maximum in the Hall coefficient) is observed (see Figs. 4.2 and 4.5 in Ref. [22]). However, in our case the temperature $T_H$ corresponding to the minimum in the $p_H$ (T) dependence is essentially higher than $T_R$, whereas in the normal SC system $T_H \approx T_R$ [22]. Note, the temperature $T_0$ is close but not equal to $T_R$, so we have three characteristic temperatures detected in the holes transport measurements: $T_R < T_0 < T_H$.

### C. Evidence of the percolation type conductivity and phase separation

Usually, at the insulating side of the percolation transition, the sample consists of metallic droplets, located in the minima of the potential relief and filled with carriers, separated by insulating barriers. In our system, this potential relief in QW reflects the long range fluctuations of potential provided by strong disorder of the Mn ions distribution in DL; this disorder is strengthened by the high content of stacking faults in the samples [18]. Since each Mn ion has the charge and local moment, there exist two components of such a relief, corresponding to the charge and spin fluctuations due to heterogeneity of the Mn positions. We presume that FM transition in DL accompanied by the Fermi level lowering inside DL (see Eq.1) serves a "trigger" to induce in QW a transition from the activation transport of holes on the percolation level to the hopping one. The switching between two types of transport is due to the decreasing of a "true" holes concentration in QW, the weakening of screening effect and the related gain of the "charge fluctuations" component of potential relief [23]. At the same time, the "spin fluctuations" component of potential relief while presumed to be small as compared to the "charge fluctuations" one, can slightly shift back the percolation transition due to the freezing of spin fluctuations in FM phase.

The small values of activation energy of the hopping conductivity and relatively high effective mobility of the hopping carriers ($\geq 10$ cm$^2$/V·s) show that metallic droplets in QW are of the mesoscopic size $D_m$ with the holes localized inside them [22, 23]. The activation energy of the hopping conductivity is determined by the droplets charge energy (energy of Coulomb blockade), $\varepsilon_c = e^2/\kappa D_m$, which is reduced due to the screening near the insulator–metal transition [24]. Following [24], we find the activation energy at hopping regime for the 2D case:

$$\varepsilon_a = \varepsilon_c \left[ 1 - \left( \frac{x_m}{x_{mc}} \right)^{1/2} \right] \approx \left( \frac{e^2}{\kappa D_m} \right) \left( \frac{D_d}{2 D_m} \right), \qquad (2)$$

where $x_m$ is the content of metallic phase and $x_{mc}$ is its critical value corresponding to the percolation threshold (for the 2D case $x_{mc} = 0.5$ [22]), $D_d$ is the thickness of dielectric barriers between the droplets (near the percolation threshold $D_d \ll D_m$). According to [23] we can estimate the $D_d$ suggesting that at $T \approx T_0$ the probability $w_a \sim \exp(-\varepsilon_a/k_BT)$ of thermal activation of holes to the percolation level becomes equal to their tunneling probability $w_t \sim \exp(-2D_d/\lambda)$ between metallic droplets. Here $\lambda = \hbar/(2m^*\varepsilon_a)^{1/2}$ is the length of the hole wave function decay under the barrier of the height $\varepsilon_a$. At $T_0 \approx 60$ K, $\varepsilon_a \approx 16$ meV and $m^* = 0.14 \cdot m_0$ [25] we obtain $D_d \approx (\hbar/2k_BT_0)\sqrt{\varepsilon_a/2m^*} \approx 6.4$ nm. Using this value of $D_d$ and the activation energy at the hopping regime $\varepsilon_a \approx 0.05$ meV one can estimate from Eq. (2) the droplet size $D_m \approx 86$ nm, i.e. in our case $D_m \gg D_d$.

The conductivity of the percolation type and large scale of electrical heterogeneity are confirmed by investigation of incoherent mesoscopic fluctuations of the non-diagonal component of a resistivity tensor $R_{xy}$ [26]. Mesoscopic fluctuations of the $R_{xy}$ reflect variation of the percolation path topology under the action of external factors (temperature, electrical and magnetic fields) and give the opportunity of direct experimental estimation of the correlation radius of the percolation cluster $L_c$ [26, 27] which determines the scale of electrical inhomogeneity. In the case of large-scale fluctuation potential (FP) for the correlation radius we have from Ref. [28]:

$$L_c = A\left(\frac{\delta\varphi}{k_BT}\right)^{\upsilon}, \qquad (3)$$

where $\nu$ is critical index of percolation theory ($\nu = 1.33$ for 2D and $\nu = 0.83$ for 3D [22]), $\delta\varphi$ is FP amplitude, $A$ is the typical FP spatial scale which in our case is equal to the droplet size ($A \approx D_m$). Equation (3) is valid at low temperatures when the $\delta\varphi$ value considerably exceeds $k_BT$ and $L_c \gg A \approx D_m$. Let us estimate $L_c$ to compare with $D_m$ value estimated above and to prove the model of long range FP.

In a heterogeneous system such as the percolation medium, noticeable voltage is always present between the Hall probes even in the absence of magnetic field. This voltage is the product of asymmetry resistance $R_a$ and longitudinal current $I_x$ through the sample. To separate the components of Hall resistance $R_H$ and $R_a$ the measurements are carried out for two opposite directions of magnetic field, and their values are calculated as $R_H(B) = [R_{xy}(B) - R_{xy}(-B)]/2$, $R_a(B) = [R_{xy}(B) + R_{xy}(-B)]/2$ (here $R_{xy} = V_y/I_x$ is transverse resistance). The $R_a$ value varies not only due to magnetoresistive effect, but also due to reconstruction of the percolation cluster at the magnetic field action [27].

Fig. 5a shows the normalized dependences of asymmetry resistance $R_a(B)$ and longitudinal resistance $R_{xx}(B)$ on the magnetic field determined at $T = 5$ K for the structures 1 and 3 with $d_s = 2$ and 5 nm. It is seen that in the case of most dielectric sample 1 the dependences $R_a(B)$ and $R_{xx}(B)$ have substantially different behavior reflecting the percolation character of conductivity. In the sample 3, regardless to similarity of $R_a(B)$ and $R_{xx}(B)$ dependences, the value $R_{xx}$ varies with the field variation strongly than $R_a$, that it also an evidence of inhomogeneous current distribution in the sample.

The observed deviation of $R_a(B)$ from $R_{xx}(B)$ can be interpreted as an effective shift of the Hall probes from each other [27]:

$$\delta L_a \approx L_p R_a(0)/R_{xx}(0)[R_a(B)R_{xx}(0)/R_a(0)R_{xx}(B) - 1] \approx L_c. \tag{4}$$

It must be emphasized that we refer to variation of the effective, rather than real, distance between the Hall probes which arises owing to reconstruction of percolation cluster and bending of the pathways of carrier transport (see insert in Fig. 5a).

In Fig. 5b the dependences $\delta L_a(B)$ normalized to the distance between potential probes $L_p$ are shown for the samples 1 (curves 1 and 1') and 3 (curves 2 and 2') at temperatures 5 and 100 K. From these data we have found for the sample 1 at 100 K that $L_c \sim 1$ μm, i.e. $L_c \gg D_m$ and corresponds to $\delta\varphi \approx 50$ meV, as follows from Eq.3. For both samples $\delta L_a/L_p$ at $T = 5$ K is significantly higher than at $T = 100$ K, as it follows from Eq. 3.

So, our structures are the percolation systems with phase separation close to insulator-metal transition (see [29] and reference therein) and contain two phases: large metallic droplets of size $D_m$ separated by narrow insulating barriers of size $D_d \ll D_m$.

### D. Hall effect in percolation two-phase systems

Magnetotransport properties of such system were theoretically studied in Ref. 30 and experimentally observed in Refs. [10, 31]. Let us discuss the properties of studied structures on the basis of the two-phase model. At the critical content of metallic phase $x_m$, $x_m = x_{mc}$ the infinite metallic cluster is formed, i.e. percolation transition occurs, which is characterized by the width $\Delta_t = |x_{mt} - x_{mc}| = (\sigma_d/\sigma_m)^m$, where $\sigma_d$, $\sigma_m$ are conductivities of insulating and metal components of the medium, respectively, and $m \approx 0.385$ for 2D case [30] (here and further $\sigma$ is the surface conductivity with the dimension $(\Omega/\square)^{-1}$). At the transition region $\Delta_t$ ($|x_m - x_{mc}| \leq \Delta_t$) the effective conductivity $\sigma(x_m)$ of the system is determined by both components [30].

For the case of a hopping conductivity inside insulating regions $\sigma_d \ll \sigma_m$, and the Hall resistance of the insulating component $R_{dH}$ is much less than that of metallic component $R_{mH}$ owing to very small value of Hall effect under hopping regime [32]. At such conditions in the vicinity of the percolation threshold, the measured Hall resistance reads [31]:

$$R_H \approx R_{mH}\left(1 - \frac{\sigma_d}{\sigma_m} A\right), \tag{5}$$

where $A = \dfrac{\sigma_d \sigma_m}{\sigma^2(x_m)}$ is a quantity of the order of unity in the $\Delta_t$ region (at insulating behaviour, $A \geq 1$). Relation (5) indicates that $R_H$ is determined by the metallic component if a sample resistance shows hopping behavior. The medium Hall mobility $\mu = \sigma \cdot (R_H/B)$ turns out to be reduced by $\approx (\sigma_d/\sigma_m)^{1/2}$ times compared to its value in metallic regions. That explains the small Hall mobility values $\mu$ for samples 1–3 compared to the value of the reference sample 4. While at intermediate temperatures when the activation transport of carriers at the percolation level dominates over the hopping one and $R_d \gg R_m$, the measured Hall resistance $R_H$ exceeds $R_{mH}$ [31]:

$$R_H \approx R_{mH}\left(1 + \dfrac{\sigma_d R_{dH}}{\sigma_m R_{mH}} A\right). \tag{6}$$

However, at relatively high temperatures, the thermal energy $(4 \cdot k_B T)$ is comparable to the activation energy at the percolation level, $R_{dH} \approx R_{mH}$, and the measured Hall resistance $R_H$ tends to $R_{mH}$. So, with the temperature growth the crossover from the hopping conductivity to the carriers transport at the percolation level passes and $R_H(T)$ should show maximum resulting in the minimum of an "apparent" concentration $p_H \propto 1/R_H(T)$, as it was observed (see Fig. 3a).

### E. Compensation of normal Hall effect by AHE, magnetization of DL and its interaction with QW

The above presented analysis could not completely explain a high value of the holes concentration $p_H$ obtained in our experiment. On the one hand, to reach the value of $p_H = (1-2)\cdot 10^{13}$ cm$^{-2}$ the potential well depth should exceed 0.4 eV. On the other hand, this $p_H$ value should take place at the transition to hopping conductivity, i.e. at $T = T_0 \approx 60$ K (see Fig. 2), contrary as it was observed at $T=T_H \approx 110$ and 80 K for samples 1 and 3, respectively (see Fig. 3a). Nevertheless, this analysis shows that the Hall effect in our system is mainly determined by metallic droplets. Therefore, the reason for a high value of $p_H$ is most likely connected not only with the shunting of the Hall effect voltage $V_H$ by the hopping conduction [22], but also with the $V_H$ compensation inside the droplets, and as a consequence with the small measured value of $R_H \propto R_{mH}$. Such compensation is possible due to the presence of normal and anomalous contributions of opposite signs to the Hall effect voltage. In this case we have

$$R_{mH} = R_{mH}^n + R_{mH}^a = \mathfrak{R}_m B - |\mathfrak{R}_s| M, \tag{7}$$

where $R_{mH}^n$ and $R_{mH}^a$ are normal and anomalous component of the Hall resistance inside metallic droplets, respectively, $\mathfrak{R}_m$ is the normal Hall effect coefficient for 2D carriers, $\mathfrak{R}_s$ is the constant of anomalous Hall effect (AHE) caused by the spin-orbit interaction of holes and their spin polarization proportional to magnetization $M$ [33]. Under a not very strong field $B$ when the moment saturation is not achieved, we have $M \cong \chi_t \cdot B$, where $\chi_t$ $(T)$ is the magnetic susceptibility in the direction perpendicular to DL. The measured Hall resistance is linear to the field, $R_{mH}/B = (\Re_m - \chi_t |\Re_s|)$, that can lead to significant difference between the holes concentration $p_H(T)$ obtained from the Hall resistance measurements and the true holes concentration $p(T)$. It was observed even a change of the sign of the Hall resistance in the helical itinerant ferromagnet MnSi where the paramagnetic component of AHE is negative, while the normal component of the Hall resistance is positive corresponding to the holes conductivity [34].

An explanation of increased values of $p_H$ in our system as a result of compensation of normal and anomalous components of the Hall resistivity is supported by the following experimental observation. The AHE in structures 1 – 3 with dielectric type of conductivity is of negative sign. Really, in Fig. 6 the magnetic field dependences of the Hall resistance $R_H$ for the sample 1 ($d_s = 2$ nm) are shown at different temperatures. The contribution of the component $R_{mH}^a$ corresponding to the ferromagnetic AHE with negative sign (the negative FM component of AHE) to the total $R_H(B)$ dependence is evident at $T= 30$ K (curve 1) and at $T= 50$ K (curve 2). We see that the absolute value of the negative ferromagnetic component of AHE diminishes with the temperature increasing: $R_{mH}^a \approx -3.5$ Ω and $R_{mH}^a \approx -2$ Ω at $T =30$ K and $T =50$ K, respectively. This component almost disappears at $T =60$ K ($R_{mH}^a \approx -0.7$ Ω) and at $T = 100$ K (curve 3) the $R_H(B)$ function linearly depends on the magnetic field that specifies a domination of the normal and/or anomalous paramagnetic components in the Hall effect. Note, that the negative sign of AHE is also observed for the reference sample 4 ($N_{Mn} \approx 0.5$ ML) with metallic type of conductivity (see inset in Fig. 6).

To explain an apparent discrepancy between the results of the conductivity and the Hall effect measurements we have to propose a self-consistent mechanism of influence of the magnetic ordering in DL on the variation of the holes concentrations $p(T)$ and $p_H (T)$ in QW. We think that this influence is due to the above discussed magnetic proximity effect. Following our assumption, hybridization between the spin polarized confinement states in DL and the holes states in QW leads to significant perceptibility of the holes in QW to the structural and magnetic disorder in DL, being stronger than $d_s$ is smaller. It is known that in DMSs such disorder causes significant fluctuations of the Coulomb and exchange potentials and becomes inevitable at high Mn content.

We presume that in our system charge and spin lateral fluctuations in DL are projected into QW due to the transverse redistribution of holes between DL and QW and quantum interference of the holes wave functions in DL and QW. In QW these fluctuations strongly modify the holes transport and form a potential relief containing metallic droplets and insulating spacers between them, with characteristic sizes $D_m$ and $D_d$, respectively. As it was mentioned above, there appear in our system three different characteristic temperatures in the holes transport in QW: $T_R \approx 46$ K, $T_0 \approx 60$ K and $T_H = 80 - 100$ K. All these temperatures depend on the spacer thickness $d_s$ and in our opinion, they reflect different stages of magnetic ordering in DL. To qualitatively describe the process of this ordering we propose the following scenario. Let us suppose that there exist in DL strong quenched disorder in Mn ions distribution and significant self-compensation of carriers provided from different (substitution or interstitial) positions of Mn ions in the GaAs lattice. Both these factors can lead to phase segregation in DL and formation of Mn enriched nanometer sized "metallic" particles ("islands") with the high holes concentration separated by nanometer sized Mn depleted "insulating" spacers ("bridges") with the low holes concentration. As a result, this phase segregation is accompanied by long scale charge and spin lateral ("*in plane*") fluctuations in DL. We presume that exchange coupling between the holes and local spins is strongly enhanced inside each "island" in DL, so that a short range FM order in it becomes favorable at the temperatures below some characteristic temperature $T_{loc} \sim 100$ K (called the temperature of "local FM transition"). Due to a smooth character of such transition it is very difficult to directly define the correct value of $T_{loc}$. But whenever, long range FM order does not appear in the system at $T < T_{loc}$, since orientations of the moments of different "islands" are chaotic. However, below $T_{loc}$ under an external magnetic field there appears an alignment of these moments along the field, but the behavior of magnetization has paramagnetic character up to the fields of about several Tesla. Such high magnetic rigidity of the system of "islands" is probably due to the effect of their blocking by a strong field of magnetic anisotropy in DL. Besides the usual (i.e. relativistic) origin, such field may have an exchange origin, for example, be going on the antiferromagnetic (AFM) order in the Mn depleted insulating regions at the periphery of FM "island" (see Ref.[6]). Apparently, strong exchange field of magnetic anisotropy is responsible for the bias of a hysteresis loop of magnetization detected in our precedent experiments [35]. Unfortunately, in the absence of detailed and trustworthy magnetic measurements we are unable to correctly reveal a type of magnetic anisotropy in DL. However, we can roughly estimate the saturation field $B_s$ in the system of uncorrelated FM "islands" suggesting that at $B = B_s$ the Zeeman energy of the "island" becomes equal to their energy of magnetic anisotropy: $M_s \cdot B_s \approx K$, where $K$ is the bulk coefficient of magnetic anisotropy, $M_s = m_{eff}(N_{Mn}/\delta)$ is the saturation magnetization of the "island", $m_{eff}$ is the effective moment per Mn atom, $\delta$ is the thickness of DL. In our case the surface concentration of Mn atoms $N_{Mn}$ is equal to $\approx 1$ ML $\approx 6 \cdot 10^{14}$ cm$^{-2}$ [15] and $m_{eff} \approx 1 \cdot \mu_B$/Mn for DL with high Mn content [24]. So, for $K \approx 10^6$ erg/cm$^3$ and $\delta \approx 2$ nm we obtain the value $B_s \approx 4$ T, which appreciably exceeds the magnetic field used in our experiments. Obviously, this value reflects the rigidity of the moments of FM "islands" over thermodynamic fluctuations and demonstrates the frozen character of these moments in the region of their existence $T < T_{loc}$ in the absence of external influence. This is why we have no significant tracks of the "local FM transition" in the $R_{xx}(T)$ measurements.

The situation becomes more complex under an influence of external magnetic field. Following our assumption, the appearance of anomalous Hall effect component in the $R_H(T)$ is due to spin polarization of droplets and their orientation in applied magnetic field perpendicular to the DL plane. For the lack of neutron scattering [36] or XAS/XMCD [37] experimental studies we are unable to establish many crucial details the magnetic structure of DL. We can only roughly associate the observed minima in the dependence Hall concentration $p_H(T)$ at $T = T_H \leq T_{loc}$ with the track of above discussed "local FM transition" in DL, where magnetization $M \cong \chi_t \cdot B$ up to $B \sim 1$ T due to a strong rigidity of FM "islands" in DL. Under these conditions (at $T \leq 100$ K) the Hall resistance starts to fall, and so "apparent" holes concentration $p_H(T)$ increases according Eq.(7) due to the $\chi_t$ growth caused by the presence of FM "islands", more strongly than $d_s$ is smaller and the holes spin polarization is higher.

Up to now we composed our scenario in terms of non-interacting FM "islands" inside DL. Let us presume, however, that at low temperatures some mechanism of exchange coupling between the moments of different FM "islands" separated by insulating "bridges" asserts oneself (for example, percolation type mechanism of indirect exchange through a narrow impurity band [38]). An effective field of this exchange overcomes the blocking effect and aligns the moments of FM "islands" thus enhancing a tendency to their parallel orientation. In frame of a simple isotropic 2D Heisenberg model we can introduce magnetic correlation length $\zeta(T)$ and write $\zeta(T) \sim D_d/(1-T_C/T)^{1/2}$ at $T >> T_C$ (paramagnetic regime) and $\zeta(T) \sim D_d/\exp(\pi T_C/T)$ at $T << T_C$ (crossover regime) where $T_C$ is the Curie temperature obtained within the mean field approach [39]. Since the true long range FM order is absent in this model at non zero temperatures, the temperature $T_C$ can be interpreted as the characteristic temperature $T_{glob}$ of "global FM transition" in the system of correlated FM "islands", where the magnetic susceptibility $\chi_t(T) \sim \zeta^2(T)$ exhibits a crossover from the power type of temperature dependence to the exponential one. By this way, for our system we can roughly identify the temperatures $T_C$ and $T_{glob}$; obviously, in frame of this conception we have to postulate that $T_{glob} << T_{loc}$. The manifestation of "global FM transition" in DL at $T=T_{glob}$ is reflected in the $R_{xx}(T)$ measurements in QW in form of a smooth peak near $T = T_C$. Strictly speaking, spontaneous magnetization does not formally exist below $T_{glob}$ and so, the ferromagnetic component of the anomalous Hall effect does not appear at $T < T_C$, while the paramagnetic component proportional to the external magnetic field exists at $T > T_C$ up to sufficiently strong fields about of 1 T. Nevertheless, if the correlation length $\zeta(T)$ exceeds the characteristic size $D_m$ of the "island" that is possible in the crossover regime, the manifestation of such (formally "short range") quasi-parallel ordering of FM "islands" in DL leads to simulate small and smooth ferromagnetic component of the anomalous Hall effect in QW(see Fig. 6). In principle, if we allow into our scenario a small Ising-like magnetic anisotropy probably existing in DL, we should expect a low temperature singularity in $\chi_t(T)$ at $T = T_0$ and appearance of a true spontaneous magnetization below $T_0$, where $T_0 \sim T_C/\ln^2(\pi^2 J/K) << T_C$ is the temperature of long range FM order, $J$ and $K$ are, respectively, isotropic and anisotropic components of an exchange integral between "islands" ($J >> K$). This anisotropy may also cause a manifestation of small and smooth ferromagnetic components of the anomalous Hall effect observed at low temperature. Unfortunately, without detailed magnetic measurements we have only uncertain hints at the tiny effects of magnetic anisotropy in our system.

## IV. CONCLUSION

Above obtained experimental results clearly demonstrate that the peculiarities of the Hall effect in GaAs/δ<Mn>/GaAs/In$_x$Ga$_{1-x}$As/GaAs structures with high Mn content are strongly connected with the magnetic proximity effect between DL and QW. They are caused by an absence of homogeneity of the DL due to the lateral and transversal disorder in the Mn ions distribution. All structures possess a dielectric character of conductivity and show the maximum or shoulder in the temperature dependence of resistance $R_{xx}(T)$ at the temperature $T_R \approx 46$ K, which is associated with the Curie temperature $T_C$ of FM transition in DL. The holes transport in QW has the percolation character and clearly demonstrates a transition to the hopping transport below $T_C$. We observe in the dependence of the Hall effect concentration $p_H(T)$ a minimum at some $T = T_H = 80$- $100$ K which value is more than less spacer thickness in QW structures. Furthermore, the increasing of $p_H$ at $T < T_H$ more strongly than $d_s$ smaller and reach high values $p_H \approx (1-2) \cdot 10^{13}$ cm$^{-2}$ at temperatures below $T_R \approx 46$ K.

We suppose that $T_H = 80$- $100$ K obtained from the temperature dependence of Hall resistance $R_H(T)$ result from a compensation of the positive sign normal Hall effect component by the negative sign anomalous Hall effect component having linear dependence on magnetic field. To explain the origin of such compensation we propose a qualitative model based on the short range magnetic ordering in 2D phase segregated system composed from Mn enriched metallic "islands" connected by Mn depleted insulating "bridges". We believe that obtained data stimulate further studies of spin polarized transport and magnetic proximity effect in FM/SC delta doped heterostructures.

ACKNOWLEDGEMENTS

We are grateful to Prof. J. Furdyna and Dr. X. Liu for their valuable comments and QW structures provided for study.

The work is partially supported by RFBR (grants 09-07-00290, 09-02-00579, 10-07-00492, 10-02-00118, 11-07-12063, 11-02-92478, 11-07-12050 and 11-02-12200).

**References**

1. K. Sato, L. Bergqvist, J. Kudrnovský, P.H. Dederichs, O. Eriksson, I. Turek, B. Sanyal, G. Bouzerar, H. Katayama-Yoshida, V.A. Dinh, T. Fukushima, H. Kizaki, and R. Zeller, Rev. Mod. Phys. **82**, 1633 (2010).
2. T. Jungwirth, Jairo Sinova, J. Mašek, J. Kučera, and A.H. MacDonald, Rev. Mod. Phys. **78**, 809 (2006).
3. Abdel F. Isakovic. *Spin Transport in Ferromagnet-Semiconductor Heterostructures*. LAP Lambert Academic Publishing (April 27, 2010), 240 p.
4. L.V. Lutsev, A.I. Stognij, and N.N. Novitskii, Phys. Rev. B **80**, 184423 (2009).
5. S. Caprara, V.V. Tugushev, P.M. Echenique, and E.V. Chulkov. Europhys. Letters, **85**, 27006 (2009).
6. V.N. Men'shov, V.V. Tugushev, S. Caprara, P.M. Echenique, and E.V. Chulkov, Phys. Rev. B **80**, 035315 (2009).
7. R.C. Myers, A.C. Gossard, and D.D. Awschalom. Phys. Rev. B, **69**, 161305 (2004).
8. S.V. Zaitsev, M.V. Dorokhin, A.S. Brichkin, O.V. Vikhrova, Yu.A. Danilov, B.N. Zvonkov, and V.D. Kulakovskii, JETP Lett. **90**, 658 (2009).
9. B.A. Aronzon, V.A. Kul'bachinskii, P.V. Gurin, A.B. Davydov, V.V. Ryl'kov, A.B. Granovskii, O.V. Vikhrova, Yu.A. Danilov, B.N. Zvonkov, Y. Horikoshi, and K. Onomitsu, JETP Lett. 85, 27 (2007).
10. B.A. Aronzon, M.A. Pankov, V.V. Rylkov, E.Z. Meilikhov, A.S. Lagutin, E.M. Pashaev, M.A. Chuev, V.V. Kvardakov, I.A. Likhachev, O.V. Vikhrova, A.V. Lashkul, E. Lähderanta, A.S. Vedeneev, and P. Kervalishvili. J. Appl. Phys., **107**, 023905 (2010).
11. R.K. Kawakami, E. Johnston-Halperin, L.F. Chen, M. Hanson, N. Guebels, J.S. Speck, A.C. Gossard, and D.D. Awschalom. Appl. Phys. Lett. **77**, 2379 (2000).
12. H. Luo, B.D. McCombe, M.H. Na, K. Mooney, F. Lehmann, X. Chen, M. Cheon, S.M. Wang, Y. Sasaki, X. Liu, and J.K. Furdyna. Physica E, **12**, 366 (2002).

13. A.M. Nazmul, T. Amemiya, Y. Shuto, S. Sugahara, and M. Tanaka. Phys. Rev. Lett. **95**, 017201 (2005); A.M. Nazmul, S. Sugahara, and M. Tanaka, Phys. Rev. B **67**, 241308 (2003).

14. U. Wurstbauer, I. Gronwald, U. Stoberl, A. Vogl, D. Schuh, D. Weiss, and W. Wegscheider. Physica E, **40**, 1563 (2008).

15. S. Caprara, V.V. Tugushev, E.V.Chulkov, Phys. Rev. B **84**, 085311 (2011).

16. R. G. Melko, R. S. Fishman, and F. A. Reboredo, Phys. Rev. B **75**,115316 (2007).

17. M.A. Chuev, I.A. Subbotin, E.M. Pashaev, V.V. Kvardakov, and B.A. Aronzon, JETP Lett. **85**, 17 (2007); M.A. Chuev, B.A. Aronzon, E.M. Pashaev, M.V. Koval'chuk, I. A. Subbotin, V.V. Rylkov, V.V. Kvardakov, P.G. Medvedev, B.N. Zvonkov, and O.V. Vikhrova, Russ. Microelectron. **37**, 73 (2008).

18. X.X. Guo, C. Herrmann, X. Kong, D. Kolovos-Vellianitis, L. Däweritz, and K.H. Ploog. J. Cryst. Growth, **278**, 655 (2005).

19. T. Wojtowicz, W.L. Lim, X. Liu, M. Dobrowolska, J.K. Furdyna, K.M. Yu, W. Walukiewicz, I. Vurgaftman, and J.R. Meyer, Appl. Phys. Lett. **83**, 4220 (2003).

20. H. Xing, S. Keller, Y-F. Wu, L. McCarthy, I.P. Smorchkova, D. Buttari, R. Coffie, D.S. Green, G. Parish, S. Heikman, L. Shen, N. Zhang, J.J. Xu, B.P. Keller, S.P. DenBaars, and U.K. Mishra, J. Phys.: Condens. Matter **13**, 7139 (2001); A. Dadgar, F. Schulze, J. Bläsing, A. Diez, A. Krost, M. Neuburger, E. Kohn, I. Daumiller, and M. Kunze, Appl. Phys. Lett. **85**, 5400 (2004).

21. K.S. Burch, D.B. Shrekenhamer, E.J. Singley J. Stephens, B.L. Sheu, R.K. Kawakami, P. Schiffer, N. Samarth, D.D. Awschalom, and D.N. Basov, Phys. Rev. Lett. **97,** 087208 (2006).

22. B.I. Shklovskii, and A.L. Efros, *Electronic Properties of Doped Semiconductors* (Springer-Verlag, New York, 1984).

23. V.A. Gergel' and R.A. Suris, Sov. Phys. JETP **48**, 95 (1978).

24. A.A. Likalter, Physica A **291**, 144 (2001).

25. J.E. Schiber, I.J. Fritz, and L.R. Dawson, Appl. Phys. Lett. **46**, 187 (1985).

26. B.A. Aronzon, V.V. Rylkov, A.S. Vedeneev, J. Leotin. Physica A **241**, 259 (1997); B. Raquet, M. Goiran, N. Negre, J. Leotin, B. Aronzon, V. Rylkov, and E. Meilikhov, Phys. Rev. B **62**, 17144 (2000); B.A. Aronzon, A.S. Vedeneev, A.A. Panferov, and V.V. Ryl'kov, Semiconductors **40**, 1055 (2006).

27. V.V. Rylkov, B.A. Aronzon, A.B. Davydov, D.Yu. Kovalev, and E.Z. Meilikhov, JETP **94**, 779 (2002).

28. B.I. Shklovskii, Sov. Phys. Semicond. **13**, 53 (1979).

29. J. Jaroszynski, T. Andrearczyk, G. Karczewski, J. Wróbel, T. Wojtowicz, D. Popović, and T. Dietl. Phys. Rev. B **76**, 045322 (2007).
30. B.I. Shklovskii, Sov. Phys. JETP **45**, 152 (1977).
31. V.V. Rylkov, B.A. Aronzon, A.S. Lagutin, V.V. Podol'skii, V.P. Lesnikov, M. Goiran, J. Galibert, B. Raquet, and J. Léotin, JETP **108**, 149 (2009).
32. T. Holstein, Phys. Rev. **124**, 1329 (1961); Yu.M. Gal'perin, E.P. German, and V.G. Karpov, Sov. Phys. JETP **72**, 193 (1992).
33. T. Dietl, in *Modern Aspects of Spin Physics*, Lecture Notes in Physics, edited by P¨otz, Walter, Fabian, Jaroslav, Hohenester, and Ulrich, Vol. 712 (Springer-Verlag, Berlin, Heidelberg, 2007), pp. 1–46.
34. M. Lee, Y. Onose, Y. Tokura, N.P. Ong. Phys. Rev. B **75**, 172403 (2007).
35. B.A. Aronzon, A.S. Lagutin, V.V. Ryl'kov, V.V. Tugushev, V.N. Men'shov, A.V. Lashkul, R. Laiho, O.V. Vikhrova, Yu.A. Danilov, and B.N. Zvonkov, JETP Lett. **87**, 164 (2008).
36. J.-H. Chung, S.J. Chung, S. Lee, B.J. Kirby, J.A. Borchers, Y.J. Cho, X. Liu, J.K. Furdyna. Phys. Rev. Lett. **101**, 237202 (2008).
37. F. Maccherozzi, M. Sperl, G. Panaccione, J. Minar, S. Polesya, H. Ebert, U. Wurstbauer, M. Hochstrasser, G. Rossi, G. Woltersdorf, W. Wegscheider, and C.H. Back, Phys. Rev. Lett. 101, 267201 (2008).
38. A. Kaminski and S. Das Sarma, Phys. Rev. B **68**, 235210 (2003).
39. D.P. Arovas and A. Auerbach, Phys. Rev. B **38**, 316 (1988); M. Takahashi, Prog. Theor. Phys. **83**, 815 (1990); D. A. Yablonskiy, Phys. Rev. B **44**, 4467 (1991).

**Figure captions**

Fig. 1. (a) GaAs/δ<Mn>/GaAs/In$_x$Ga$_{1-x}$As/GaAs heterostructure with the Mn δ-layer remote from In$_x$Ga$_{1-x}$As quantum well. (b) X-ray rocking curve for the sample 2 ($d_s$ = 3 nm, $N_{Mn}$ = 1 ML). The dashed lines are experimental data, and the solid curves are calculation results. The points indicate the contribution of noncoherent diffuse scattering. (c) The profiles of the lattice parameter $\Delta a/a$ variation for the sample 2 ($a$ is the lattice parameter for GaAs). The dashed lines indicate errors.

Fig. 2. Temperature dependence of normalized resistance for samples 1-4 with $d_s$ = 2 nm, $N_{Mn}$ = 1 ML (curve 1); $d_s$ = 3 nm, $N_{Mn}$ = 1 ML (curve 2); $d_s$ = 5 nm, $N_{Mn}$ = 1 ML (curve 3); $d_s$ = 3 nm, $N_{Mn}$ = 0.5 ML (curve 4). The curve number corresponds to the sample number. The inset illustrates the temperature dependences of resistance for samples 1 and 3 in log$R_{xx}$ vs 1/$T$ coordinates.

Fig. 3. Temperature dependences of the Hall effect concentration (a) and mobility (b) of holes for samples 1 with $d_s$ = 2 nm, $N_{Mn}$ = 1 ML (curve 1) and 3 with $d_s$ = 5 nm, $N_{Mn}$ = 1 ML (curve 2).

Fig.4. Temperature dependence of the hole concentration and mobility for the reference sample 4 ($d_s$ = 3 nm, $N_{Mn}$ = 0.5 ML).

Fig.5. (a) Normalized dependences of the asymmetry resistance $R_a$ and longitudinal resistance $R_{xx}$ on magnetic field for samples 1 ($d_s$ = 2 nm, $N_{Mn}$ = 1 ML) and 3 ($d_s$ = 5 nm, $N_{Mn}$ = 1 ML) at 5 K. The curves 1 and 1' are the $R_a(B)$ and $R_{xx}(B)$ dependences for sample 1, respectively; curves 2 and 2' are the $R_a(B)$ and $R_{xx}(B)$ dependences for sample 3. The insert illustrates occurrence of mesoscopic fluctuations of the non-diagonal component of resistivity tensor $R_{xy}$. (b) Dependences of the relative "displacement" (spurious) of Hall probes on magnetic field for sample 1 (curves 1, 1') and 3 (curves 2, 2') at 5 and 100 K.

Fig.6. Dependence of the Hall resistance $R_H$ for the sample 1 ($d_s$ = 2 nm, $N_{Mn}$ = 1 ML) on magnetic field at different temperatures: 30 K (curve 1), 50K (curve 2), and 100K (curve 3). Inset: the anomalous component of Hall resistance $R_H$ versus the magnetic field for the reference sample 4 ($d_s$ = 3 nm, $N_{Mn}$ = 0.5 ML) at 32 K.

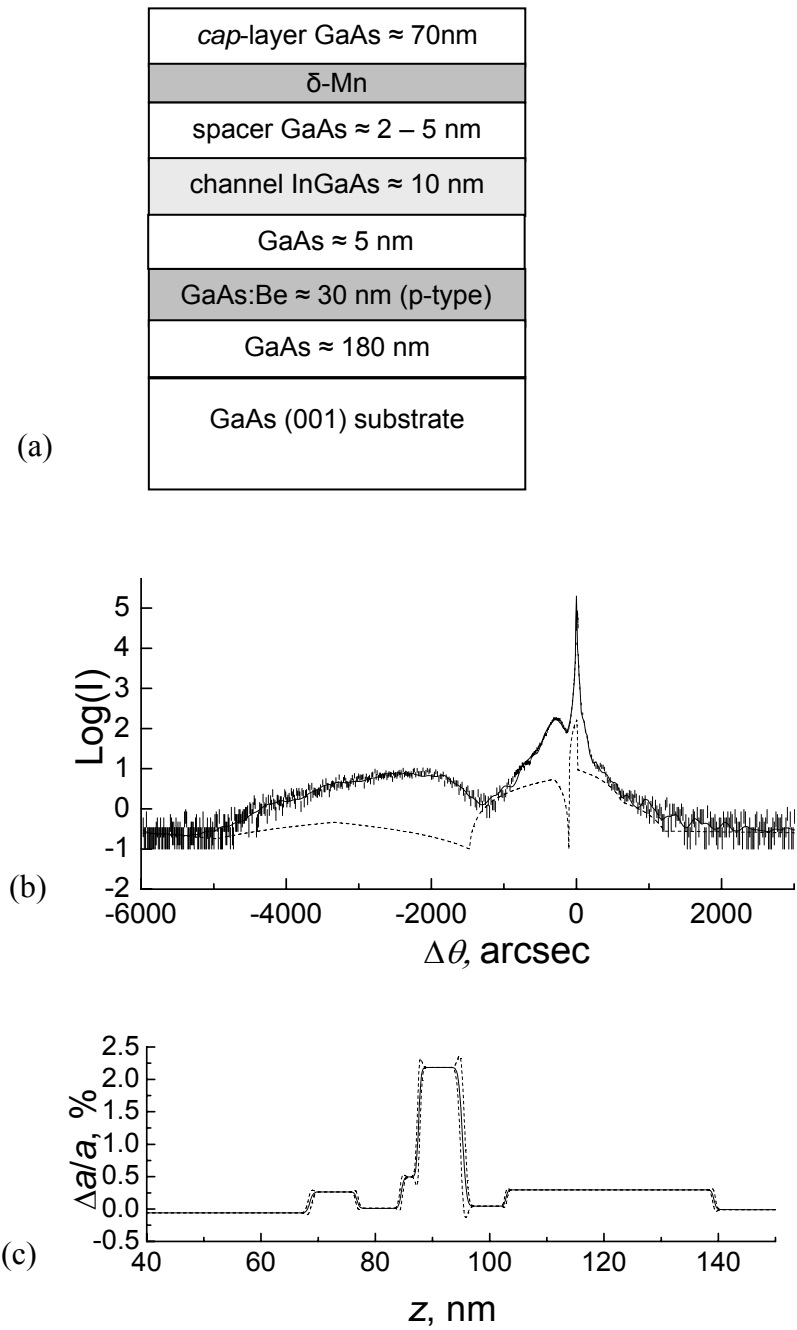

Fig.1.

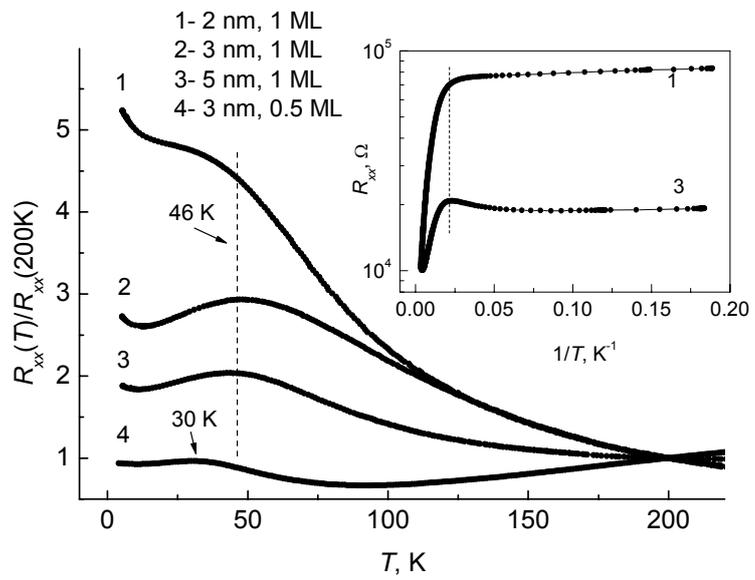

Fig.2.

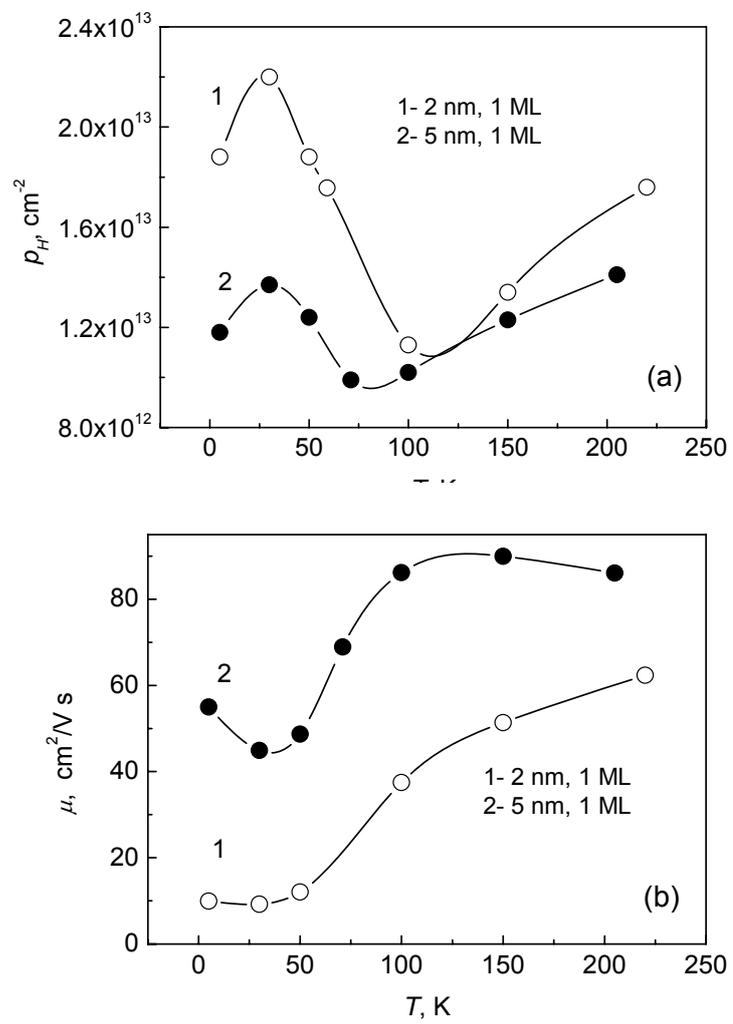

Fig.3.

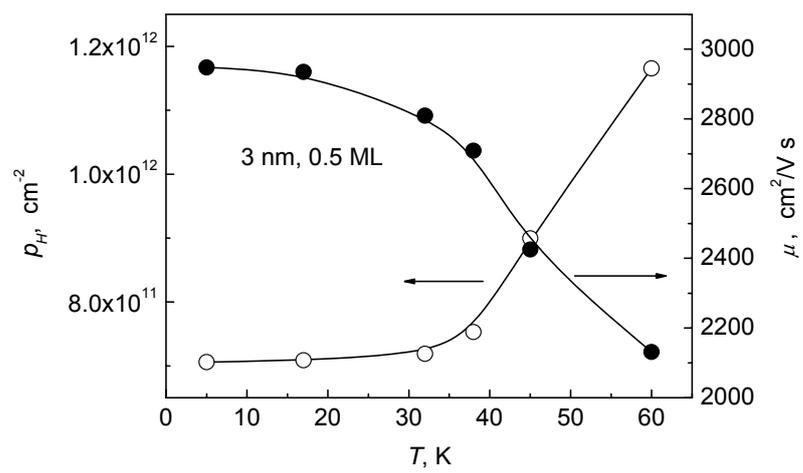

Fig.4.

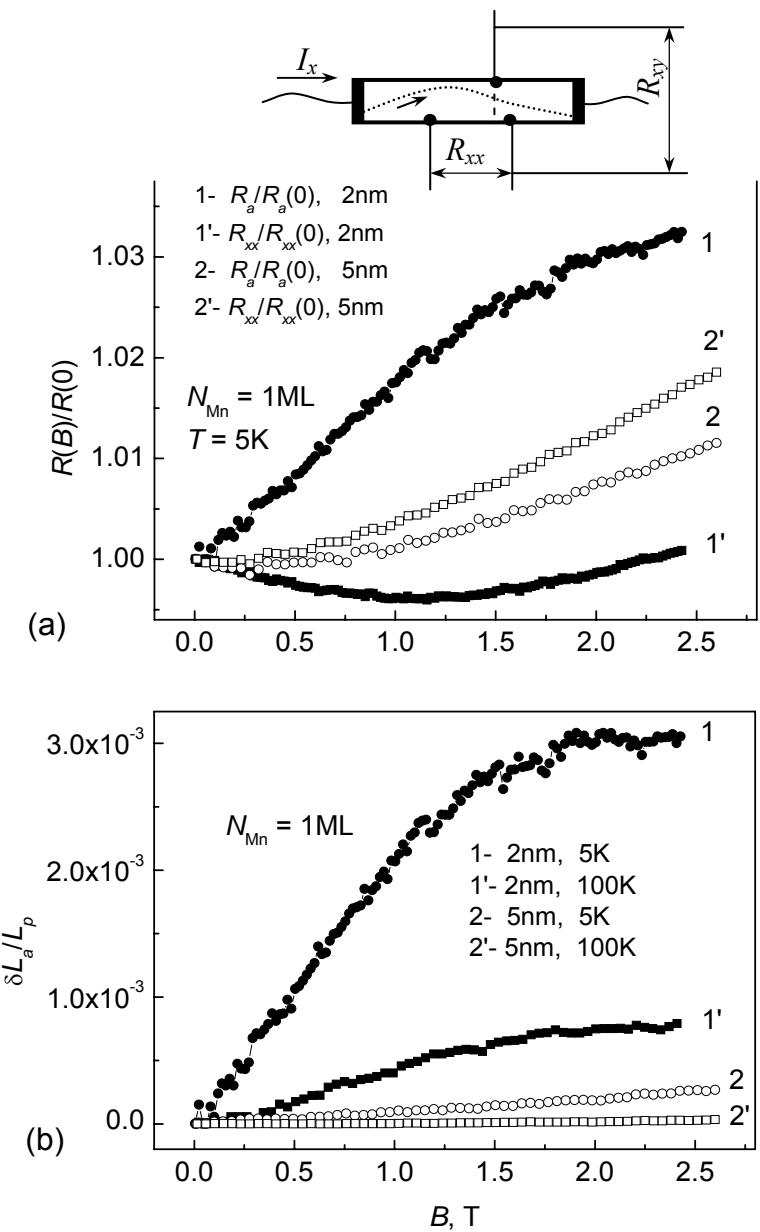

Fig.5.

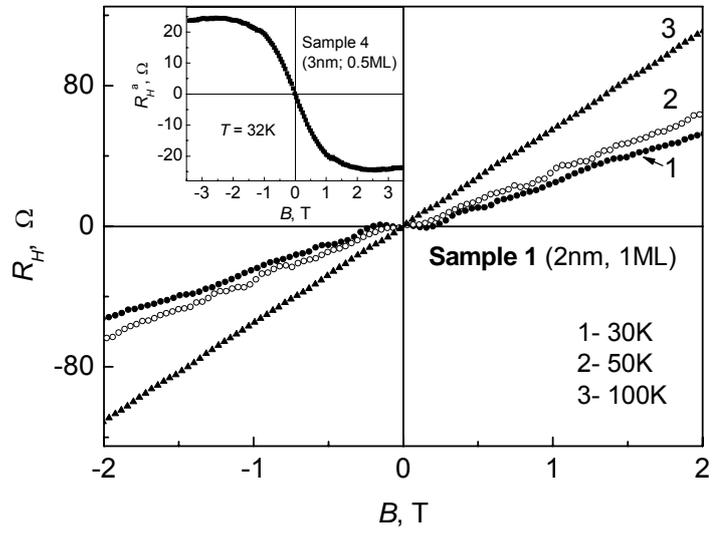

Fig.6.